\documentclass[11pt]{iopart}

\usepackage{iopams}  

\usepackage{color}
\usepackage{hyperref}
\usepackage[usenames,dvipsnames,svgnames,table]{xcolor}
\usepackage{ulem}

\usepackage{mathrsfs} 

\newcommand{\msc}{\mathscr}
\newcommand{\mc}{\mathcal}
\newcommand{\f}[2]{\frac{#1}{#2}}
\newcommand{\kron}[2]{\delta^{#1}_{\phantom{#1}#2}}
\newcommand{\chr}[2]{\Gamma^{#1}_{#2}}
\newcommand{\p}{\phantom}
\newcommand{\ud}[2]{^{#1}_{\phantom{#1 } #2}}

\begin{document}

\paper[Linear gravitational memory]{The sky pattern of the linearized gravitational memory effect}

\author{Thomas M\"adler$^1$\footnote{Email:tm513@cam.ac.uk} and Jeffrey Winicour$^{2,3}$}

\address{
${}^{1}$  Institute of Astronomy, University of Cambridge, Madingley Road, Cambridge, CB3 0HA, UK \\
${}^{2}$ Department of Physics and Astronomy \\
        University of Pittsburgh, Pittsburgh, PA 15260, USA\\
${}^{3}$ Max-Planck-Institut f\" ur
         Gravitationsphysik, Albert-Einstein-Institut, \\
	 14476 Golm, Germany \\
	}

\begin{abstract}
The gravitational memory effect leads to a net displacement
in the relative positions of test particles. This memory is related to the change in the
strain of the gravitational radiation field between infinite past and infinite future retarded times.
There are three known sources of the memory effect: (i) the loss of energy to future null infinity by massless fields
or particles, (ii) the ejection of massive particles to infinity from a bound system and (iii) homogeneous, source-free
gravitational waves. In the context of linearized theory, we show that asymptotic conditions controlling these
known sources of the gravitational memory effect rule out any
other possible sources with physically reasonable stress-energy tensors.
Except for the source-free gravitational waves, the two other known sources produce gravitational memory
with E-mode radiation strain, characterized by a certain curl-free sky pattern of their polarization.
Thus our results show that the only known source of B-mode gravitational memory is of primordial origin,
corresponding in the linearized theory to a homogeneous wave entering from past null infinity.

\end{abstract}



\section{Introduction}

The gravitational memory effect produces a net displacement between test particles
after the passage of a gravitational wave. The
effect was initially recognized in linearized gravity where the radiation memory is produced
by an exploding system of massive particles which escape to infinity~\cite{zeld}.
Independently such burst memory was found as a zero frequency mode of gravitational radiation 
in studies of the collision of relativistic point particles~\cite{Smarr} and  in studies of astrophysical
gravitational wave emission~\cite{BontzPrice}. A nonlinear form of radiation memory was
discovered  by  Christodoulou~\cite{Christ_mem}, which  was then related~\cite{thorne,will}  to the flow
of gravitons to null infinity. Subsequent studies showed that even in linearized gravitational theory the
analogous energy transport to future null infinity $\mathcal{I}^+$
by massless particles or fields (neutrinos, electromagnetic
radiation, etc.) also produce gravitational radiation memory~\cite{Bieri_null,Bieri_neutrino,tolish};
so that rather than {\it nonlinear}  memory this is now referred to as {\it null} memory.
Despite the stunning recent observation of gravitational  waves by the
LIGO-Virgo consortium~\cite{LIGO},
the detection of gravitational memory is more demanding due to the current insensitivity
of the LIGO detectors,  as well as pulsar timing arrays~\cite{NANOGrav},
to the long rise-time of typical memory signals~\cite{Favata}.

An overlooked aspect of gravitational radiation memory has been its global sky pattern,
which can be decomposed into E  and B  modes similar to the decomposition of electromagnetic
radiation, as explained in Sec.~\ref{sec:lin}. Recently, it has been shown that both the gravitational
burst memory due to ejected particles and the null
memory are purely E-mode~\cite{globalemmem}, where it was also pointed out that B-mode
gravitational memory could be produced by a homogeneous, source-free gravitational wave,
as we explicitly demonstrate in Sec.~\ref{sec:homwaves}. 

In this paper, we refine and elaborate the global properties of gravitational radiation memory in the context of 
linearized gravitational theory. We show that if burst memory (or its
time reversed counterpart consisting of the capture of particles incident from
infinity) and homogeneous wave memory
are eliminated by a weak asymptotic stationarity condition then null memory is the
only possible form of gravitational memory arising from matter with a
physically reasonable stress-energy tensor. Except for homogeneous wave memory, the
other two known sources (burst and null) produce E-mode gravitational memory.
Thus our results imply that B-mode memory is of primordial origin, corresponding
in the linearized theory to a homogeneous wave entering from past null infinity.
These results apply to non-compact matter sources with isotropic stress, scalar fields
and electromagnetic fields. 

The gravitational memory effect is an asymptotic feature which arises from the difference in the radiation
strain at ${\mathcal I}^+$ between infinite past and infinite future retarded times. As we discuss in
Sec.~\ref{sec:lin}, there is a connection between the gravitational
memory effect and the supertranslation freedom in the
Bondi-Metzner-Sachs (BMS) asymptotic symmetry group~\cite{Sachs_BMS}.
For this reason, we base our work upon a linearized version of
the null hypersurface formulation of the Einstein equations pioneered
by Bondi et al~\cite{bondi} and by Sachs~\cite{sachs}.
The aspects of the Bondi-Sachs formulation related to gravitational wave memory 
are reviewed in Sec.~\ref{sec:bs}. While this is a metric based
approach, we concentrate on gauge invariant quantities.
For an alternative covariant formalism in terms of the Weyl tensor see~\cite{Bieri_cov}.

In Sec.~\ref{sec:asymp}, we present the underlying assumptions regarding asymptotic
flatness and the asymptotic conditions on the matter stress-energy tensor.
In Sec.~\ref{sec:astat_metric}, we discuss a weak asymptotic stationarity
condition which controls incoming radiation fields and rules out two of the
known sources of the gravitational memory effect: the burst memory and
the homogeneous wave memory. The weak asymptotic stationarity
condition requires that the leading $r^{-5}$ coefficient of the
Newman-Penrose curvature component $\Psi_0$ be stationary in the retarded time limits
$u=\pm \infty$. Our main results for E-mode memory are given in Sec.~\ref{sec:mem_E}; and those for
B-mode memory, in Sec.~\ref{sec:mem_B}.
The role of matter fields  is discussed in Sec.~\ref{sec:mem_matter}. 

Our approach is based upon a previous study of the global properties of electromagnetic
radiation memory~\cite{globalemmem}, which showed that only the known
current distributions which carry charge to infinity can produce
E-mode memory~\cite{bieri_couter} and that
B-mode electromagnetic 
radiation memory cannot be be produced by a physically
realistic charge-current distribution.
In the course of this work, it was realized that the version of asymptotic stationarity condition
assumed in~\cite{globalemmem} was unnecessarily strong. 
In~\ref{app:em}, we revisit the electromagnetic memory results under a weaker condition
of asymptotic stationarity analogous to the gravitational version presented here. 
 
As pointed out in~\cite{bieri_couter}, the electromagnetic memory effect is more dramatic than the
gravitational effect. Electromagnetic waves produced by the ejection
of a charged particle give rise to a net momentum kick to test charges in the radiation zone, as opposed
to the displacement of test masses in the gravitational effect.
Other recent work on aspects of the memory effect addresses its role in
cosmological effects~{\cite{bieri_dS, cosmo_tail, cosmo_tolish,cosmo_BMS}}, in binary neutron star mergers~\cite{Bieri_em},
in higher dimensional theories~\cite{mem_higherD},
in properties of angular momentum~ \cite{spin_mem,mem_angular},
in the supertranslation freedom~\cite{mem_soft_theorem}
and in the black hole information paradox \cite{bhmem2015,soft_info}.

We use geometric units in which the Einsteinian gravitational constant is $\kappa=8\pi$.
The signature of the metric is $+2$ and we use the conventions of~\cite{MTW} for the curvature. 
For reference, a list of the linearized Christoffel symbols is provided in~\ref{app:christ}


\section{Radiation fields on the celestial sphere, geometrical framework}
\label{sec:lin}

In linearized theory, gravitational radiation is described by a trace free strain tensor
on the celestial sphere $x^A=(\theta,\phi)$, 
\begin{equation}
      \sigma_{AB}(u,x^A) , \quad q^{AB}  \sigma_{AB} =0,
\end{equation}
where $q_{AB}dx^A dx^B=d\theta^2+\sin^2 \theta d\phi^2$ is the standard unit sphere metric
and $u$ is the retarded time.
The radiation memory is determined by the change in the strain between infinite future and past
retarded time,
\begin{equation}\label{def_memory}
       \Delta  \sigma_{AB}: = \sigma_{AB}(u=\infty,\theta,\phi) -  \sigma_{AB}(u=-\infty,\theta,\phi).
\end{equation}    
This produces a net displacement in the relative angular position of distant test particles
\begin{equation}
      \Delta (  x_2^A -  x_1^A) =q^{AB}  \Delta \sigma_{BC}(  x_2^C -  x_1^C) ,
\end{equation}
where we use the notation  $\Delta F(x^C):=F(u,x^C)|_{u=\infty}-F(u,x^C)|_{u=-\infty}$. 

Similarly, the electromagnetic radiation
pattern at future null infinity ${\mathcal I}^+$ is characterised by the angular components
of the (rescaled) electric field $E_A(u,x^B)$.
The E and B mode classification results
from decomposing the electric field in terms of a gradient and the dual of a gradient,
\begin{equation}\label{dual_decomp}
           E_A = D_A \Phi_{[e]} + \epsilon_{CA} D^C \Phi_{[b]},
\end{equation}
where $D_A$ is the covariant derivative and $\epsilon_{AB}$ 
is the antisymmetric surface area tensor on the unit sphere, respectively. The real scalar
fields  $\Phi_{[e]}$ and $\Phi_{[b]}$ correspond to the E-mode and B-mode, respectively.

A compact way to describe this decomposition is in terms of
a complex polarisation vector $q_A$ satisfying 
\begin{equation}
          q _{AB} = q_{(A} \bar q_{B)}, \quad \epsilon_{AB} =i q_{[A} \bar q_{B]},
            \quad q^A \bar q_A =2, \quad q^A q_A =0,
\end{equation}
e.g. $q_A = (1, i \sin \theta)$ for the standard form of the unit sphere metric and where
$(AB)$ and $[AB]$ denote the standard notation for symmetrization and antisymmetrization
of indices.
Then the electromagnetic radiation is represented by the spin-weight 1 field
\begin{equation}
          q^A E_A = q^A (D_A \Phi_{[e]} +\epsilon^C_{\p{C}A}D_C \Phi_{[b]}) = \eth(\Phi_{[e]}+i \Phi_{[b]}),
\end{equation}
where $\eth$ is the Newman-Penrose spin-weight raising operator~\cite{BMS2,penrin}.

The analogous decomposition of the gravitational radiation field can be made by first noting
that the strain can be represented in terms of a displacement vector field $\xi^A$  by
\begin{equation}\label{def_strain_sigma_AB}
    \sigma_{AB}=\frac{1}{2}(D_A \xi_B +D_B \xi_A) 
    -\frac {1}{4} q_{AB} D^C \xi_C .
\end{equation}
Given two real scalar fields $\Sigma_{[e]}$ and $\Sigma_{[b]}$ that form the complex
scalar field $\Sigma:=\Sigma_{[e]}+i\Sigma_{[b]}$, the decomposition
\begin{equation}
     \xi_A = D_A \Sigma_{[e]} + \epsilon_{BA} D^B \Sigma_{[b]}
\end{equation} 
leads to the spin-weight 2 representation of the strain,
\begin{equation}\label{def_rad_mem_spin0}
     \sigma:=q^A q^B \sigma_{AB}=q^A q^B D_A D_B(\Sigma_{[e]}+i\Sigma_{[b]}) 
     =\eth^2(\Sigma_{[e]}+i\Sigma_{[b]})=\eth^2\Sigma\;.
     \label{eq:sigma0}
\end{equation}
In spin-weight terminology, $\Sigma_{[e]}$ and $\Sigma_{[b]}$ represent the 
``electric'' and ``magnetic'' parts of the strain, corresponding to the E and B
radiation modes. Here $\Sigma$ is the {spin-weight 0}
potential generating the {spin-weight 2} field $\sigma$ via the  spin-weight raising operator $\eth$,
according to (\ref{eq:sigma0}). As we will make extensive use of the spin-weight
calculus,  if $J_{A_1\dots A_n}$ is a symmetric trace free tensor  field on the sphere
and $\eth^n J= q^{A_1}\dots q^{A_n}J_{A_1\dots A_n}$, 
we denote the real part of the corresponding spin-weight 0 potential $J$
by $J_{[e]}$ and its imaginary part by $J_{[b]}$.

In the electromagnetic case,
the $\vec E$\ and $\vec B$ radiation fields have equal magnitude and are orthogonal
in direction. Thus the electromagnetic radiation pattern and its memory could be equally
well described in terms its $\vec B$ field.
Similarly, in the gravitational case, the radiation fields corresponding to $ \vec E$ and $\vec B$, in
an inertial frame picked out by a time-like vector $T^a$, arise
from the components of the Weyl tensor and its dual, 
\begin{equation}
    {\bf E} = T^a T^b q^A q^B C_{aAbB}, 
    \quad {\bf B} =- \frac{1}{2} T^a T^b  q^A q^B \epsilon_{aAcC}{C^{cC}}_{bB}\;,
\end{equation}
where $\epsilon_{0123}=1$ is the four dimensional Levi Civita tensor.
Here it follows from the Petrov type $N$ of the radiation field that the spin-weight 2
fields ${\bf E}$ and ${\bf B}$
are equal in magnitude and oriented at $45^o$, i.e.  ${\bf E} \rightarrow {\bf B}$
under the spin rotation $q^A \rightarrow e^{i\pi/4} q^A$.
The leading coefficients of the electric and magnetic parts of the Weyl tensor in an expansion at 
${\mathcal I}^+$  are related to the strain according to
\begin{equation}\label{E_B_sigma}
  E(u,x^A) =         {\partial^2_u} \sigma(u,x^A)\;, \qquad
    B(u,x^A) =         i{\partial^2_u} \sigma(u,x^A)\;.
\end{equation}
For our purposes, it will be useful to deal with the strain but the Weyl tensor
representation has the advantage of being gauge invariant. The strain has gauge freedom
\begin{equation}
\label{supertranslation}
\sigma \rightarrow \sigma+\eth^2 \alpha
\end{equation}
under a retarded time transformation $u\rightarrow u+\alpha(x^A)$, which corresponds
to the supertranslation freedom in the Bondi-Metzner-Sachs asymptotic symmetry
group~\cite{Sachs_BMS,BMS2}. Thus, on comparison with \eref{supertranslation}, the E-mode
component of the strain can be gauged away during any stationary epoch. However,
the E-mode of the memory \eref{def_memory} is gauge invariant and can be considered to
be a supertranslation shift between the two preferred gauges for the strain
picked out at $u=\pm \infty$. 
It is curious, and perhaps of some deeper significance, that in the electromagnetic case the 
E-mode radiation memory is  also related to a gauge shift, in that case
with respect to the vector potential~\cite{globalemmem} (see~\ref{app:em}).


\section{Linearized Bondi-Sachs Framework}
\label{sec:bs}

It will be useful to refer to three separate coordinate systems. A Cartesian inertial system
$(t,x^i)=(t,x,y,z)$, the associated spherical coordinate system $(t,r,x^A)=(t,r,\theta,\phi)$,
$r^2 =x^2+y^2+z^2$,  and the associated outgoing null spherical coordinate system
$x^\alpha=(u,r,x^A)$, with retarded time $u=t-r$ and vertices of the null cones at $r=0$.
In these retarded null coordinates, the Bondi-Sachs line element is
\begin{equation}
     ds^2= g_{uu} du^2 +2 g_{ur} du dr + 2 g_{uA} du dx^A + r^2h_{AB} dx^A dx^B,
\end{equation}
where the use of an areal radial coordinate $r$ implies that $\det (h_{AB})= \det (q_{AB})$.
In the linearized approximation off the Minkowski background metric $\eta_{ab}$,
\numparts
\begin{eqnarray}
g_{uu} &\approx &-1 -2\beta -W,\\
    g_{ur}  &\approx &-1 -2\beta , \\
     g_{uA}  &\approx &-r^2 q_{AB}U^B ,\\
      h_{AB} &\approx& q_{AB} +J_{AB},\qquad q^{AB}J_{AB} =0\;.
\end{eqnarray}
\endnumparts
In the following we treat $(\beta,W,U^A,J_{AB})$ as linearized quantities and neglect higher order terms.
In this approximation, the non-zero contravariant metric components are   
\numparts
\begin{eqnarray}
g^{ur}&=&-1+2\beta , \\
 g^{rr} &=&1-2\beta +W ,\\
    g^{rA}&=& -U^A,\\
     g^{AB} &=&r^{-2}(q^{AB}-J^{AB}),\qquad J^{AB}: =q^{AC}q^{BD}J_{CD}.
\end{eqnarray}
\endnumparts

Instead of working with the metric variables, we introduce spin-weighted fields
and express covariant {derivative} $D_A$ with respect to the unit sphere
metric in terms of the $\eth$ operator \cite{BMS2,eth}. For example,
we represent the traceless symmetric tensor $J_{AB}$ by the pure
spin-weight $2$ field $\mc{J}=q^{A}q^B J_{AB}$ and write
\begin{equation}\label{def_eth}
       \eth \mc{J} = q^{A}q^B q^C D_C J_{AB}\;,\quad
          \bar \eth \mc{J} = q^{A}q^B \bar q^C D_C J_{AB}\;.
\end{equation}
A spin-weight $s$ field $\mc{F}(x^a)$ satisfies the commutation relation
$[\bar\eth, \eth] \mc{F}= 2s \mc{F}$.

The  linearized metric is determined by $\beta$, $W$, the spin-weight 1 field
$\mathcal {U} = q_A U^A$ and the spin-weight 2 field $ \mathcal {J} = q^Aq^B J_{AB}$.
In the same way, we represent the components  $\rho_{ab}$ of the reduced
stress-energy tensor of the standard matter stress-energy tensor  $T_{ab}$,
\begin{equation}
\label{red_Tab}
\rho_{ab}:=T_{ab}-\f{1}{2}\eta_{ab}T\ud{c}{c},
\end{equation}
by the spin-weighted fields 
\begin{equation}
\mathcal{N} = q^A {\rho}_{uA}\,,\quad
\mathcal{P} = q^A {\rho}_{rA}\,,\quad
\mathcal{S} = q^A q^B {\rho}_{AB},
\end{equation}
and the spin weight 0 field $S_{0} =q^A \bar q^B {\rho}_{AB}$.

A spin-weight $s$ field $\mc{F}(x^a)$  can be expanded
in terms of spin-weighted harmonics $_sY_{lm}(x^A)$ $(l\ge s)$,
\begin{equation}
\mc{F}(x^a) = \sum_{l=s}^\infty \sum_{m=-l}^l f_{l m}(u,r)\,_sY_{l m}(x^A).
\end{equation}
Here the ${}_s Y_{lm}$ are generated
by applying $\eth^s$ to the standard spherical harmonics $Y_{lm}(x^A)$.
\footnote{The $s<0$ case is obtained by exchanging $\eth$ by $\bar\eth$.}.
By this procedure, we can introduce a complex spin-weight 0 potential $F$
such that $\mc{F}=\eth^s F$. 
Hereafter, we denote any spin-weight $s\neq 0$ quantity by a scripted font, e.g. $\mc{F}$,
and the its spin-weight 0 potential with the corresponding Roman font, e.g $F$. This
procedure is not a priori unique since $\eth^s Y_{lm} =0$ for $l < s$. We remove this
ambiguity by imposing the convention that for a spin-weight s field $\mc{F}$ its corresponding
potential $F$ contains no harmonics with $l <s$.

Application of this notation to the metric and matter variables
leads to the spin-weight 0 potentials
\begin{equation}
\label{s0_fields}
\mathcal{U} = \eth U\,,\quad
\mathcal{J} = \eth^2 J\,,\quad
\mathcal{N} = \eth N\,,\quad
\mathcal{P} = \eth P\,,\quad
\mathcal{S} = \eth^2 S .
\end{equation}
Note that due to the above convention, $U$, $P$ and $N$ have no $l=0$ modes
(which otherwise would not be of physical significance). Similarly, $J$ and $S$ have
no $l=0$ and no $l=1$ modes.
This notation leads to a natural decomposition of the linearized  field equations
into their E and B-modes corresponding, respectively,
to their real and imaginary parts, as in the decomposition of
the strain \eref{def_rad_mem_spin0}.


\subsection{ Einstein Equations}

We express the linearized Einstein equations in the form
\begin{equation}
\label{EE}
R_{ab} = \kappa \rho_{ab}\;.
\end{equation}
Following the formalism in~\cite{newt,nullinf},
they decompose into the hypersurface equations
\numparts
\begin{eqnarray}
\fl \partial_r \beta = \f{\kappa}{4}r\rho_{rr}\;,\\
\fl\partial_r \left(r^4 \partial_r\mc{U}\right) 
=
   2r^4\partial_r\left(\f{1}{r^2}\eth \beta\right)
   \label{eq:ethU_lin} 
  -\f{1}{2} r^2\overline \eth \partial_r\mc{J}
  + 2\kappa r^2 \mc{P}\;, \\
\fl 2\partial_r(rW)
 =
 4\beta
    +\f{1}{4}(\eth^2 \overline{\mc{J}}+\overline \eth^2 \mc{J})
  -2\overline\eth\eth \beta
  +\f{1}{2r^2}\partial_r \Big[r^4( \eth \overline{\mc{U}} +  \overline \eth \mc{U})\Big]
  -\kappa S_0\;, 
\end{eqnarray}
\endnumparts
the complex evolution equation
\begin{eqnarray}
\fl   r\partial_r\partial_u(r\mc{J})
&=& 
\f{1}{2}\partial_r\left(r^2\partial_r\mc{J}\right)
-\eth \partial_r\Big(r^2\mc{U}\Big)
     +2 \eth^2\beta
     +\kappa \mc{S}     \;,\label{eq:ev_ethU_lin} 
\end{eqnarray}
and the supplementary conditions
\numparts
\begin{eqnarray}
\fl \kappa (-\rho_{uu}+\rho_{ur}) \! = \!\!
-\partial_u\bigg[
  	 \f{W}{r} -\f{2}{r}\beta
   -\f{1}{2}(\eth \overline{ \mathcal{U} }+\overline \eth \mathcal{U})\bigg]
     - \f{1}{2r^2}\overline \eth\eth W 
      -\f{1}{4r^2}\partial_r\Big[r^2(\overline \eth \mathcal{U} + \eth \overline {\mathcal{U}})\Big] ,
      \\
\fl \kappa \mc{N}  =  \partial_u \Big(\f{1}{4}\overline\eth\mc{ J}   - \eth\beta
    -\f{1}{2} r^2 \partial_r \mc{U}
    \Big)
    +    \f{1}{2} \eth\partial_rW
     +\f{1}{2}\partial^2_r(r^2\mc{U}) 
     +\f{1}{4}(\eth \overline \eth \mc{U} - \eth^2\overline {\mc{U}}) \;.
\end{eqnarray}
\endnumparts
The supplementary conditions can be interpreted as conservation equations \cite{tam,Goldberg}.
It follows from the Bianchi identities  and the matter conservation laws 
that if the hypersurface and evolution equations
are satisfied then the supplementary conditions are automatically satisfied if they
are satisfied on a worldtube $r=R(u,x^A)$ (or if the vertex world line $r=0$ is nonsingular).

Rewritten in terms of the spin-weight 0 fields $(U,J,N,P,S)$, these linearized equations 
reduce to the hypersurface equations
\numparts
\begin{eqnarray}
\fl   &{}&  \partial_r\beta = \f{\kappa}{4}r\rho_{rr}\;,\label{hyp_beta_s0}\\
 \fl &{}&  \partial_r \left(r^4 \partial_r U\right) =
   2r^4\partial_r\left(\f{1}{r^2} \beta\right)
  -\f{1}{2} (\overline \eth \eth +2)r^2\partial_r J
  + 2\kappa r^2 P\label{hyp_U_s0} \;,\\
\fl   &{}& 2\partial_r(rW) =4\beta
      +\frac{1}{2}\overline \eth\eth (\overline \eth \eth +2) J_{[e]}
  -2\overline\eth\eth \beta
  +\f{1}{r^2}\partial_r \Big(r^4\overline \eth \eth\,U_{[e]}\Big)
  -\kappa S_0\;,
  \label{hyp_W_s0}
\end{eqnarray}  
\endnumparts
the evolution equation
\begin{eqnarray}
\fl   r\partial_r\partial_u(rJ)  &=& 
   \f{1}{2}\partial_r\left(r^2\partial_rJ\right)
- \partial_r(r^2U)
       +2 \beta
     +\kappa S, \label{ev_s0}
\end{eqnarray}
and the supplementary  equations
\numparts
\begin{eqnarray}
\label{supp_rr_u_s0}
\fl \kappa (-\rho_{uu}+\rho_{ur})
=
 -\partial_u\left(
  	 \f{W}{r} -\f{2}{r}\beta
   - \overline{\eth}\eth  U_{[e]}\right)
     - \f{1}{2r^2}\overline \eth\eth W
      -\f{1 }{2r^2}\partial_r \Big(r^2 \overline{\eth}\eth U_{[e]}\Big)\;,\\
\fl \kappa N = 
\partial_u \Big[\f{1}{4}(\overline\eth\eth +2)J   - \beta
    -\f{1}{2} r^2 \partial_r U
    \Big]
    +   \f{1}{2} \partial_rW
     +\f{1}{2}\partial^2_r(r^2U)
     +\f{i}{2}\overline \eth \eth U_{[b]}\;.
     \label{supp_rr_A_s0}
\end{eqnarray}
\endnumparts

Given the gravitational data $J$ on an initial null hypersurface $u_0$
and the matter data $\rho_{rr}$, $P$, $S$ and $S_0$,
the hypersurface and evolution equations consist of radial
differential equations which can be integrated 
at retarded time $u_0$ to determine $\beta, U, W, \partial_u J$,
in that sequential order. The corresponding integration constants are a mixture
of physical properties of the system, e.g. the mass and angular momentum aspects, 
and gauge information. As explained
in Sec.~\ref{sec:asymp_metric}, it is possible to use the gauge freedom to set
$$\beta(u,r,x^A)|_{r=\infty} =U(u,r,x^A)|_{r=\infty} =J(u,r,x^A)|_{r=\infty}=0$$.

The Newman--Penrose Weyl component $\Psi_0$ \cite{NPa,NPb} is given in terms of the
gravitational data by\footnote{The factor $1/4$ in (\ref{psi0_defs0}) arises because of the normalization
of the null vectors and $Q^a$.} 
  \begin{equation}
\label{psi0_defs0}
\Psi_0 = -\f{1}{4} C_{abcd}K^aQ^bK^cQ^d= \f{1}{8r^2}\partial_r(r^2\partial_r \eth^2 J) ,
\end{equation}
where $C_{abcd}$ is the Weyl tensor. Here $Q_a = (0,0,rq_A)$ and 
\begin{equation}
K_a = -\nabla_a u \;,\qquad  N_a = -\nabla_a v 
\end{equation}
are the null vectors associated with the retarded time $u=t-r$ and the
advanced time $v=t+r$, with normalization $K^aN_a= -2$.


\section{Asymptotic behavior}
\label{sec:asymp}

We are interested in systems which are asymptotically flat at future null infinity \cite{null-infinity}.
In particular, we impose the peeling property which requires that the leading terms
in the asymptotic expansion of the Bondi-Sachs metric can be expressed as a
power series in $1/r$. Thus, to the required power of $1/r$, we assume
that a field  $F(x^a)$ describing the metric or matter has an expansion
\begin{equation}
\label{asyp_exp}
F(x^a) = F_{[0]}(u,x^A) +  \f{F_{[1]}(u,x^A)}{r}+\f{F_{[2]}(u,x^A)}{r^2}+\f{F_{[3]}(u,x^A)}{r^4}+....,
\end{equation}
where the coefficients functions $F_{[n]}(u,x^A)$ are evaluated at $\mc{I}^+$.
In particular, $F(u,r=\infty, x^A) = F_{[0]}(u,x^A)$.

We make the following physical assumptions on the matter at null infinity:
\begin{enumerate}
  \item The total  energy, momentum and angular momentum of the matter, 
  and  their related fluxes are finite.
 \item The dominant  energy condition holds.
 \item The matter stress-energy tensor satisfies the local conservation law
  $\nabla_a T\ud{a}{b}=0$.
\end{enumerate}
These matter conditions are complemented by asymptotic flatness and gauge
conditions on the perturbed metric.


\subsection{Matter asymptotics}
\label{sec:asymp_matter}

The most general asymptotic conditions on a stress-energy tensor $T_{ab}$ 
are that the total energy, momentum and angular momentum, and their fluxes
are finite. These conditions indirectly apply to the stresses by requiring a
dominant energy condition. 

The condition that the total energy and momentum
content of the matter be finite, i.e.  $\int_0^\infty \xi^a {T_a}^u r^2 dr$  
converges for every background translational  Killing vector $\xi^a$, implies 
\numparts
 \begin{equation}
\label{asymp:energy}
T_u^u=O(r^{-4})\;,\;\;
T_r^u=O(r^{-4})\;,\;\;
T_A^u=O(r^{-3})\;,
\end{equation}
Index symmetry of $T_{ab}$ then implies 
\begin{equation}
T_r^r=O(r^{-4})\;,\;\;T_r^A=O(r^{-5}) .
\end{equation}

The condition that the matter
transport finite energy-momentum and angular momentum to
infinity, i.e. that the flux $\xi^a {T_a}^r r^2$ be finite at ${\mathcal I}^+$
for each Minkowski background space-time translational or rotational
Killing vector $\xi^a$, requires
\begin{equation}
\label{fin_flux_cond}
T_u^r=O(r^{-2})\;,\qquad T_A^r=O(r^{-2})\;, \qquad T_u^A =O(r^{-4})\; ,
\end{equation}
where we have again used index symmetry to constrain $T_u^A$.  
Furthermore, the requirement that the time integrated fluxes
$\int_{-\infty}^\infty \xi^a {T_a}^r r^2du$ also be finite  gives stronger conditions
in the infinite past and future (see also App. of \cite{Bieri_cov}),
\begin{equation}
\label{fin_int_flux}
T_u^r=O(r^{-3})\;,\qquad T_A^r=O(r^{-3})\;, \qquad T_u^A =O(r^{-5})\;, \qquad u=\pm \infty\; .
\end{equation}

The dominant energy condition requires that for any observer
the velocity of the matter is alway smaller than the velocity of light.
Specifically, for every unit timelike vector $\tau^a$, with $T_{ab}\tau^a\tau^b\ge0$
and $T_{ab}\tau^a$ a non-spacelike vector, an equivalent formulation of the
dominant energy condition
is that in any orthonormal basis $(\tau^a, e^a_{(i)}),\;i=1,2,3$,
 $T_{ab}\tau^a\tau^b\ge |T_{(i)(j)}|$ for each $i,j$~\cite{HE}. This implies
 conditions on the stress components.
The vector $\tau^a = (r^{-1/4}, \f{1}{2} r^{1/4},0,0)$, with norm $\tau^a \tau_a =-(1+r^{-1/2}) $,
approaches a unit timelike vector as $r\rightarrow\infty$. Together with \eref{fin_int_flux} this
implies $T_{ab}\tau^a\tau^b = O(r^{-3.5})$ at $u=\pm\infty$. 
Since we require that the leading terms in an asymptotic expansion be integer powers of $1/r$,
for an orthonormal base associated with $\tau^a$ at $\mc{I}^+$
the dominant energy condition implies 
\begin{equation}
\label{strong_gen}
T\ud{A}{B} = O(r^{-4})\;,\;\;
u=\pm\infty\;.
\end{equation}
\endnumparts

The relations \eref{asymp:energy}-\eref{strong_gen} manifest the most general
asymptotic behaviour of the $T_{ab}$ under the given assumptions. 
Given these results, the leading terms in the conservation laws
$\nabla_b T_a^b=0$ imply at $u=\pm\infty $
\begin{equation}\label{cons_law_T}
    0=    \partial_u T_{u[4]}^u - T_{u[3]}^r 
       = \partial_u T_{r[4]}^u
    =    \partial_u T_{A[3]}^u .
\end{equation}
In terms of the reduced tensor $\rho_{ab}$,  the asymptotic behaviour  \eref{asymp:energy}-\eref{strong_gen}
correspond to 
\numparts
\begin{equation}
\label{cond_rho_ab_1}
\rho_{uu} = O(r^{-2})\;, \;\; \rho_{ur} = O(r^{-3})\;,\;\; \rho_{rr} = O(r^{-4})\;,\;\;
\end{equation}
and for the matter spin-weight 0 potentials
\begin{equation}
\label{mat_s0_cond}
N=O(r^{-2})\;,\;\;
P=O(r^{-3})\;.
\end{equation}
Further, the finiteness of the time-integrated flux 
along with the dominant energy condition require at $u=\pm\infty$
\begin{equation}
\label{fin_int_flux_rhouu_N}
\rho_{uu} = O(r^{-3})\;\;,\;\;
N = O(r^{-3})\;\;,\;\;
S_{0} = O(r^{-2})\;,\;\;
S= O(r^{-2})\;.
\end{equation}
The conservations laws \eref{cons_law_T} in terms of $\rho_{ab}$ yield at $u=\pm\infty$
\begin{equation}
\label{asympt_mat_du}
0= \partial_u S_{0[2]} -2\rho_{uu[3]} 
= \partial_u \rho_{rr[4]}
 = \partial_u P_{[3]} \;.
\end{equation}
\endnumparts

\subsection{Spacetime asymptotics
}\label{sec:asymp_metric}

The asymptotic behavior of the matter tensor in the previous section ensures that the radial integration of
the hypersurface and evolution equations is convergent as $r \rightarrow \infty$.
Some of the integration constants involved in these radial integrals represent pure
gauge effects and others represent physical quantities such as the mass and
angular momentum aspects. Radiation
memory is an asymptotic effect whose treatment is simplified
by taking advantage of the asymptotic gauge freedom.

For a given gauge transformation $\xi^a$ (i.e. a linearized {diffeomorphism}),
the linearized metric undergoes the gauge transformation \cite{MTW}
\begin{equation}
 \delta_{\underline{\xi}} g^{ab} = g^{ac}\partial_c \xi^b +g^{cb}\partial_c \xi^a
-\xi^c \partial_c g^{ab} .
\end{equation}
This gauge freedom is subject to the Bondi-Sachs coordinate
conditions $\delta_{\underline{\xi}} g^{uu}=0$, $\delta_{\underline{\xi}} g^{uA}=0$ and
$g_{AB}\delta_{\underline{\xi}} g^{AB}=0$ \cite{Sachs_BMS,bondi},
which leads to the functional dependence
\begin{equation}
        \xi^u = \alpha(u,x^B)\;,\qquad 
      \xi^A=f^A(u,x^B)-\frac{1}{r}D^A \alpha\;,\quad
     \xi^r= -\frac{r}{2}D_B \xi^B. \qquad
\end{equation}
By setting $q_A\xi^A=\xi$ and $q_Af^A=\eth f$, in terms of a complex
scalar field $f(u,x^A)$, we obtain
\begin{equation}
\xi=\eth f - r^{-1}\eth \alpha\;,\quad
    \xi^r =-\f{r}{2}\overline \eth \eth f_{[e]} +\f{1}{2} \overline \eth \eth \alpha  \;.
\end{equation}
This gives rise to the following gauge freedom in the linearized metric variables:
\begin{equation}
 2 \delta_{\underline{\xi}} \beta = - \partial_u \alpha
                     - \partial_r \xi^r
                   =  - \partial_u \alpha
                     + \frac{1}{2}  \eth \bar \eth f_{[e]} \;,
                     \qquad 
\end{equation}
\begin{equation}
   \delta_{\underline{\xi}} \mc{U} = (\partial_u - \partial_r) \xi
            - {1 \over r^2} \eth \xi^r
            =\partial_u \eth f 
            +\f{1}{2r}\eth(\overline \eth \eth f_{[e]}-2\partial_u\alpha)
            +\f{1}{2r^2}\eth(\overline \eth \eth+2)\alpha ,
\end{equation}
\begin{equation}\label{gauge_J}
    \delta_{\underline{\xi}} \mc{J }= -\eth^2 f + { 1 \over r} \eth^2 \alpha,
\end{equation}
and
\begin{equation}
\delta_{\underline{\xi}} W = -\partial_u(2\xi^r+\xi^u)+\partial_r\xi^r 
 = r\partial_u\overline \eth \eth f_{[e]} 
 -\partial_u(\overline \eth \eth +1 )\alpha 
 -\f{1}{2}\overline \eth \eth f_{[e]}
 +\f{1}{2r}\overline \eth \eth \alpha .
\end{equation}
This gauge freedom allows us to fix the integration constants at ${\mathcal I}^+$.
 In particular,
\begin{equation}
    \lim_{r\rightarrow\infty}   \delta_{\underline{\xi}} \mc{U} = \eth \partial_u f (u,x^A) 
\end{equation}
so that we can use the $u$-dependence in the $f (u,x^A)$
gauge freedom to set $\mc{U}_{[0]} =0$, in the notation of \eref{asyp_exp}.
Next,
\begin{equation}
    \lim_{r\rightarrow\infty}      \delta_{\underline{\xi}} \beta  =-\partial_u \alpha(u,x^A)  +\frac{1}{2}\eth\bar\eth f_{[e]}(x^A)
\end{equation}
so that we can use the $u$-dependence in the $\alpha(u,x^A)$ gauge freedom
to set $\beta_{[0]} =0$.
The remaining gauge freedom is time independent and is determined by
$\alpha(x^A)$ and $f(x^A)$. At some fixed retarded time $u=u_0$ we have
\begin{equation}\label{fix_J_at_scri}
      \lim_{r\rightarrow\infty}    \delta_{\underline{\xi}} \mc{J} = -\eth^2 f (x^A), 
\end{equation}
which allows us to use the gauge freedom in $f(x^A)$ to set $ \mc{J}_{[0]}(u_0, x^A)  =0$.

In summary, after using gauge freedom, the spin-weight 0 metric fields satisfy 
\begin{equation}
      \beta_{[0]}(u,x^A) =0, \quad U_{[0]}(u,x^A) =0, \quad J_{[0]} (u_0,x^A)=0,
\end{equation}
with the remaining gauge freedom determined by $\alpha(x^A)$.

Now, with these asymptotic conditions on the metric variables
along with the asymptotic properties of the matter variables,
consider the radial integration of the hypersurface and evolution equations.
Given the asymptotic matter conditions $\rho_{rr}=O(r^{-4})$ and  $P=O(r^{-3})$,
the integration of the $\beta$-hypersurface equation \eref{hyp_beta_s0} and
$U$-hypersurface equation \eref{hyp_U_s0} imply $\beta=O(r^{-2})$ and 
$U=O(r^{-2})$.  As a result, the $J$-evolution equation \eref{ev_s0}, together
with $S=O(r^{-2})$, implies
\begin{equation}
   \partial_u J_{[0]}(u, x^A) =0 \qquad 
\end{equation}
so that, with the initial gauge condition  $J_{[0]}(u_0,x^A) =0$,
we have $J_{[0]}(u,x^A) =0$ at all times.
Thus, we can use the gauge freedom to set
\begin{equation}
\label{metric_gauge}
     \beta=\beta_{[2]}r^{-2}  +..., \quad  U=U_{[2]}r^{-2} +..., 
\quad  J=J_{[1]}r^{-1} +... ,
\end{equation}
where the remainder terms are of higher order. 
Choosing  $\xi_a = -\nabla_a u$ in the formula for the
radiation strain (\ref{def_strain_sigma_AB})  gives
\footnote{For the choice $\xi_A = -\nabla_A u $ in (\ref{def_strain_sigma_AB})
the  norm $|\sigma| $ equals  the fractional length change $\delta L/L $ measured
by a gravitational wave detector.  }
\begin{equation}\label{asyp_strain_calc}
          \sigma(u,x^A)=\frac{r}{2} q^A q^B J_{AB}(u,r,x^A)|_{r=\infty}=\frac{1}{2}\eth^2 J_{[1]}.
\end{equation}
Together with (\ref{def_memory}) and (\ref{def_rad_mem_spin0}),
(\ref{asyp_strain_calc}) shows that the spin-weight 0 potential $\Delta \Sigma$ for the radiation memory
$\Delta \sigma$ is
\begin{equation}
\label{rad_memory}
\Delta \Sigma = \f{1}{2}\Delta J_{[1]}.
\end{equation}
The remaining gauge freedom  (\ref{fix_J_at_scri}) of $J$ is  according to \eref{gauge_J}
\begin{equation}
      \delta_{\underline{\xi}}  J_{[1]}(u,x^A) = \alpha(x^A), 
\end{equation} 
so that $J_{[1]}$ may be gauged to $0$ in either  the limit $u=\infty$
or the limit $u=-\infty$, but the difference $\Delta J_{[1]}$ is gauge invariant.


\section{Weak Asymptotic Stationarity}
\label{sec:astat_metric}

The weak asymptotic stationarity condition (see \eref{ASC}) is key to our analysis
of the sky pattern of the memory effect. It serves to control incoming radiation and
eliminates two of the three known sources of radiation memory: burst memory and
homogeneous wave memory. The third known source, i.e. null memory, depends
upon a non-zero integrated energy flux to ${\mathcal I}^+$.

\subsection{Notation and useful formulae}

Here we present some useful relations between the
flat space Cartesian coordinates $x^a=(t, x^i)$, together with the associated orthonormal tetrad
\begin{equation}
T_a = -\partial_a t\;,\quad
X_a = \partial_a x\;,\quad
Y_a = \partial_a y\;,\quad
Z_a = \partial_a z\; ,
\end{equation}
and their spherical analogues.
Associated with the spherical coordinate $r$ 
are the unit vector
$r_a =\partial_a r$ and its the second partial derivative $r_{ab} = \partial_a\partial_b r$,
with the spatial components
\begin{equation}
   r_i=\frac{x_i}{r}=(\sin\theta\cos\phi,\sin\theta\sin\phi,\cos\theta),
       \quad r_{ij}=
            \frac{\delta_{ij}}{r}-\frac{x_i x_j}{r^3}\; ,
\end{equation}
so that  $x^a = (t, rr^i)$.
With this notation,  the second derivative of a function $F(t,r)$ is given by
\begin{eqnarray}
\label{G_ab}
\fl \partial_a\partial_b F= \Big(\partial_t^2 F + \f{\partial_r F}{r}\Big)T_a T_b 
	-2(\partial_t\partial_r F)T_{(a}r_{b)}
	+\Big(\partial_r^2 F - \f{\partial_r F}{r}\Big)r_a r_b 
	+\Big(\f{\partial_r F}{r}\Big)\eta_{ab} .
\end{eqnarray}

The  vector $Q^a$ with $Q^a T_a = Q^a r_a=0$
and angular components $Q^A = q^A/r$
has Cartesian spatial components
\begin{equation}
       Q^i =\f{1}{r}\f{\partial x^i}{\partial x^A}q^A = (\cos\theta\cos\phi-i\sin\phi,
             \cos\theta\sin\phi+i\cos\phi,-\sin\theta)\;, 
\end{equation}
which satisfy
\begin{equation}
Q_a x^a = 0\;\;,\;\;
Q^a \partial_a Q^b = \f{\cot\theta}{r}Q^b \;,\;\;
Q^i r_{ij} = \f{1}{r}Q_j\;,\;\;
Q^a\partial_a x^A = q^A/r\;\; 
\end{equation}
and the useful relations
\numparts
\begin{equation}
        (Q^x)^2+ (Q^y)^2 = -\sin^2\theta, \quad
\end{equation}
\begin{equation}
     Q^x r_y-Q^y r_x= -i\sin\theta, \quad
\end{equation}
\begin{equation}
     Q^x r_x+Q^y r_y= \sin\theta\cos \theta, \quad
\end{equation}
\begin{equation}
     Q^x r_y +Q^y r_x  =\sin\theta \bigg[
         2\cos\theta\cos\phi\sin\phi +i(\cos^2\phi-\sin^2\phi) \bigg]\;,
\end{equation}
\begin{equation}
     Q_a Q_b(X^a Y^c-Y^a X^c) (X^bY^d-Y^bX^d)r_c r_d
       =-\sin^2\theta\;,
\end{equation}       
\begin{equation}
     Q_a Q_b(X^a Y^i-Y^a X^i) (X^bY^j-Y^bX^j)\delta_{ij}
       =-\sin^2\theta\;.
\end{equation} 
\endnumparts

\subsection{Weak Asymptotic Stationarity Condition}\label{sec:wasc}
\subsubsection{Boosted linearized Schwarzschild solution}
\label{sec:astat_mono}

A particle at rest with mass $m$ gives rise to the linearized Schwarzschild metric 
\begin{equation}
g_{ab} = -\Big(1-\f{ 2m }{r}\Big)T_aT_b + \Big(1+\f{ 2m }{r}\Big)(X_aX_b+Y_aY_b+Z_aZ_b).
\end{equation}
Burst memory is produced by a particle initially at rest which is later
ejected with escape velocity $V$.
For a boost in the $z$-direction
with four velocity $v^a = \Gamma(1,0,0,V)$, with  $\Gamma = (1-V^2)^{-1/2}$, 
the resulting memory is
\begin{equation}
\label{mem_bss}
\Delta \sigma =  -\f{ 2m \Gamma V^2 \sin^2\theta }{( V\cos\theta-1)} \;\;.
\end{equation}

In order to see how burst memory can be eliminated in a gauge invariant way
by a weak asymptotic stationarity condition, consider the corresponding
curvature tensor, which for the Schwarzschild solution is determined
by the electric part $E_{ab}$ of the Weyl tensor. In the rest frame,
 $E_{ab}r^ar^b =  2m /r^3$ and  $E_{ab}Q^a\bar Q^b = - 2m/r^3$
 so that
\begin{equation}
E_{ab} = \f{ m }{r^3} \Big(3r_a r_b -\eta_{ab}- T_aT_b\Big)\;,
\end{equation}
where $T^a$ is the 4-velocity of the particle.
The corresponding Weyl tensor is
\begin{equation}
\label{weyl_rest}
    C_{abcd}= \f{ 2m }{r^3}\Big[12\,  r_{[a}T_{b]}r_{[c}T_{d]} 
     +6\Big(  T_{[a}\eta_{b][c}T_{d]}	
	    -r_{[a}\eta_{b][c}r_{d]}	\Big)
	-2 \eta_{a[c}\eta_{d]b}\Big]\;.
\end{equation}
For a boosted particle with 4-velocity $T^a \rightarrow v^a$
and $r$ and $r_a$ substituted by the Lorentz covariant expressions
\begin{equation}
\label{R_Ra}
r^2\rightarrow R^2 = x^ax_a+ (v_ax^a)^2
\;\;,\qquad
r_a\rightarrow R_a =\f{1}{R}[x_a + (v_bx^b)v_a] ,
\end{equation}
we  obtain
\begin{equation}
\label{general_Weyl}
C_{abcd}  = \f{ 2m }{R^3}\Big[12  R_{[a}v_{b]}R_{[c}v_{c]} 	         
	    +6 \Big(
	   v_{[a}\eta_{b][c}v_{d]}	
	    -R_{[a}\eta_{b][c}R_{d]}	
	\Big)
 	-2 \eta_{a[c}\eta_{d]b}
	\Big]\;. 
\end{equation}
The corresponding Weyl component $\Psi_0$ is
\begin{eqnarray}
\Psi_0 & = &
- \f{3m  }{2R^5} \Big[(x_a K^a)(v_bQ^b)- (v_a K^a)(x_b Q^b)\Big]^2 
\end{eqnarray} 
Since $x_a Q^a=0$ and $x_aK^a=-u$, we find
for a  boost in the z-direction with velocity $V$
\begin{equation}
\label{Psi0_bss}
\Psi_0 
  =- \f{ 3m  \Gamma^2 u^2 V^2 \sin^2\theta}{2R^5} 
  =- \f{3 m}{2}  \f{V^2 \sin^2\theta}{\Gamma^{3}(1+V\cos\theta)^{5/2}}\f{u^2}{r^5} +O(r^{-6}) .
\end{equation}

For a particle at initially at rest, $\Psi_0=0$ and $\Psi_0$ satisfies
(\ref{Psi0_bss}) after it is ejected. Thus the ejection and
the associated burst memory can be eliminated by
requiring the {\it weak stationarity condition} 
\begin{equation}
\label{ASC}
    0=\Delta \lim_{r\rightarrow \infty} r^5 \partial_u^2 \Psi_0.
\end{equation}
This also rules out the memory effect due to the time reversed process of
the capture of particles incident from infinity. 

Note that  \eref{ASC} does not restrict higher order boosted multipoles, whose Weyl curvature $\Psi_0$
falls off faster than $1/r^5$. However, as we show next, this weak asymptotic stationarity condition
does rule out the memory effect due to source-free waves.
Expressed in terms of the metric variable $J$ by means of \eref{psi0_defs0},
weak asymptotic stationarity (\ref{ASC})  restricts the metric  coefficient $J_{[3]}$ according to
\begin{equation}
\label{ASC2}
  0 = \Delta \partial^2_u J_{[3]}\;.
\end{equation}
In particular, this places no restriction on the time dependence of
the radiation field $J_{[1]}$, and therefore it is not a direct restriction on the
gravitational memory $\Delta \Sigma$.

\subsubsection{Source-free gravitational waves}
\label{sec:homwaves}


Here we consider the standard formalism \cite{MTW} for linearized metric
perturbations $h_{ ab } $ of a Minkowski background metric,
\begin{equation}
\label{standard_lin_pert}
g_{ab} = \eta_{ab}+h_{ab}, 
\quad g^{ab} = \eta^{ab}-h^{ab},
\quad \det(g_{ab}) = -1 + h^a_a
\end{equation}
where indices are raised and lowered with the Minkowski metric.
The densitized version of the perturbation  is given by
\begin{equation}
   \sqrt{-g} g_{ab} = \eta_{ab}+\gamma_{ab}\;,
\end{equation}
where  
\begin{equation}
\gamma_{ab}=h_{ab}-\frac{1}{2}\eta_{ab} h^c_c .
\end{equation}
We adopt the harmonic gauge condition $\partial_b \gamma^{ab} =0$,
in which the  source-free linearized  Einstein equations take the simple form
$ \eta^{ab} \partial_a\partial_b \gamma_{cd}=0. $

We construct source free linearized waves in the
harmonic gauge by using the gravitational analogue
of a Hertz potential $H^{acbd}$~\cite{ Boardmann,SachsBergmann}, which has the symmetries
\begin{equation}
  H^{acbd}=  H^{[ac]bd}=H^{ac[bd]}
      =H^{bdac}
\end{equation}
and satisfies the flat space wave equation
$ \partial^e \partial_e H^{acbd}= 0$.
As a result, the densitized metric perturbation
\begin{equation}
      \gamma^{ab} =\partial_c \partial_d H^{acbd}
\end{equation}
satisfies the linearized Einstein equations in the harmonic gauge.

Source-free, ingoing-outgoing gravitational waves can be generated from the potential
\begin{equation}\label{def_gen_H_abcd}
    H^{acbd} =K^{acbd}\frac{f(t-r)-f(t+r)}{r}\; ,\;\; \partial_{e}K^{{abcd}}=0\;\;,
\end{equation}
which gives rise to the perturbation
\begin{equation}\label{def_gamma_ab_via_f}
      \gamma^{ab} =K^{acbd}
    \partial_b \partial_c \frac{f(t-r)-f(t+r)}{r}\; .    
\end{equation}
According to \eref{asyp_strain_calc}, the strain of the radiation field on ${\mathcal I}^+$ is given by
\begin{eqnarray}
   \sigma &=& \lim_{r\rightarrow \infty} \frac{1}{2} r Q_a Q_b \gamma^{ab} .
\end{eqnarray}
For these vacuum solutions of the field equations $R_{abcd}=C_{abcd}$ and the
Weyl component  $\Psi_0$ is
\begin{equation}
  \Psi_0 = -\frac{1}{4}R_{abcd} K^a Q^b K^c Q^d =
  - \frac{1}{8}(2\gamma_{ab,cd}-\gamma_{ac,bd}-\gamma_{bd,ac})K^a Q^b K^c Q^d\; , 
  \label{eq:psi0}
\end{equation}
where the individual terms are determined by the tetrad components of
$\gamma_{ab}$ according to
\numparts
\begin{eqnarray}
 \fl      \gamma_{ab,cd} Q^a Q^b K^c K^d  =
 K^{c}\partial_{c}\Big [K^{d}\partial_{d}(\gamma_{ab} Q^a Q^b)\Big ] \;\;,  \label{eq:pieces_1st}\\
  \fl     \gamma_{ab,cd} Q^a K^b Q^c K^d =
  K^{c}\partial_{c}\Big [ Q^{d}\partial_{d}\Big(\frac{1}{\sin\theta}\gamma_{ab} Q^a K^b\Big)  \sin\theta
           -\frac{1}{r}\gamma_{ab} Q^a Q^b\Big] \;, \\
  \fl     \gamma_{ab,cd} K^a K^b Q^c Q^d  =
     Q^c\partial_{c} \Big[\frac{Q^{d}\partial_d(\gamma_{ab} K^a K^b)}{\sin\theta}  \Big]\sin\theta
  -\frac{4 \sin\theta }{r}Q^{c}\partial_{c}\Big(\frac{\gamma_{ab} Q^a K^b}{\sin\theta}\Big)
     +\frac{2}{r^2}\gamma_{ab} Q^a Q^b .
     \nonumber\\
           \label{eq:pieces_last} 
\end{eqnarray}
\endnumparts
Considering  $\gamma_{ab} Q^a Q^b$, $\gamma_{ab} K^a Q^b$ and $\gamma_{ab} K^a K^b$
to be functions of $(u, r, \theta,\phi)$, we rewrite \eref{eq:pieces_1st} -\eref{eq:pieces_last} as
\numparts
\begin{eqnarray}
    \fl   \gamma_{ab,cd} Q^a Q^b K^c K^d = \partial_r^2(\gamma_{ab} Q^a Q^b) ,
    \label{eq:piecesmod_1st}\\
    \fl   \gamma_{ab,cd} Q^a K^b Q^c K^d 
    = 
       \partial_r \Bigg[\f{\sin\theta}{r} q^A\partial_A\Big(\frac{\gamma_{ab} Q^a K^b}{\sin\theta}\Big) \Bigg]
           -\partial_r\Big[\frac{\gamma_{ab} Q^a Q^b}{r}\Big] \;\;, \\
    \fl  \gamma_{ab,cd} K^a K^b Q^c Q^d =\f{\sin\theta}{r^2} q^A \partial_A\bigg[\frac{q^A\partial_A(\gamma_{ab} K^a K^b)}{\sin\theta}  
       	-\frac{4\gamma_{ab} Q^a K^b}{\sin\theta}\bigg]
         +\frac{2}{r^2}\gamma_{ab} Q^a Q^b .
           \label{eq:piecesmod_last}
\end{eqnarray}
\endnumparts
For our purpose, it suffices to consider the perturbation \eref{def_gamma_ab_via_f}
determined by
\begin{eqnarray} \label{eq:C}
f(\tau) = 
\left\{
\begin{array}{ccc}
\f{1}{2}C \tau^2& : &T<\tau \\
F(\tau)& : &0\le \tau \le T \\
0& : &\tau<0 
\end{array}\right.
\end{eqnarray}
where $F(\tau)$ is chosen to make the solution smooth.
In particular,
\begin{eqnarray}
       f''(u) &=&C  , \quad  f''(v) = C  ,\quad \;:\;\;u > T \nonumber \\
       f''(u) &=& 0 , \quad  f''(v) = C,\quad \;\;:\;\; u <0 \;\mbox{and} \; u+2r > T  .
\end{eqnarray}


\subsubsection{Quadrupole B-mode gravitational wave memory}
\label{sec:qbw}

The choice 
\begin{equation}
  K^{abcd}= \f{1}{4}\Big(T^{[a}Z^{b]}X^{[c}Y^{d]} + X^{[a}Y^{b]}T^{[c}Z^{d]}\Big)
       \label{eq:qmgrwave}
\end{equation}
in \eref{def_gamma_ab_via_f}, along with the help of \eref{G_ab},
gives rise to the purely B-mode quadrupole perturbation with components
\numparts
\begin{eqnarray}
  Q^\alpha Q^\beta \gamma_{\alpha\beta} &=&2i \sin^2\theta
             \bigg( \frac{f''(u)+f''(u+2r)}{r}+\frac{f'(u)-f'(u+2r)}{r^2} \bigg ), 
  \label{QQ_g_mag} \\
  K^\alpha Q^\beta \gamma_{\alpha\beta} &=& i \cos\theta\sin\theta
   \bigg(\frac{-2f''(u+2r)}{r}+\frac{2f'(u)+4f'(u+2r)} {r^2} 
   \nonumber\\&&\qquad\quad
   + \frac{3f(u)-3f(u+2r)} {r^3} \ \bigg),
  \label{QK_g_mag}  \\
  K^\alpha K^\beta \gamma_{\alpha\beta} &=&  0 ,
      \label{KK_g_mag} 
\end{eqnarray}
\endnumparts
corresponding to spin-weight $(l=2,m=0)$ spherical harmonics.
This perturbation gives rise to the  B-mode radiation strain
\begin{equation}
   \sigma_{[b]} (u,x^A) = - i\sin^2\theta \lim_{r\rightarrow \infty}  \Big[f''(u) + f''(u+2r)\Big]\;.
\end{equation}
 The waveform (\ref{eq:C})  then leads to the non-zero radiation memory
\begin{equation}\label{mem_mag_pure}
     \Delta \sigma_{[b]} =  -i C \sin^2\theta .
\end{equation}
Thus homogeneous waves with non-zero B-mode radiation memory exist.

In order to show that weak asymptotic stationarity rules out these waves
we calculate the Weyl component $\Psi_0$. Substitution of \eref{QQ_g_mag} - \eref{KK_g_mag}
into \eref{eq:piecesmod_1st} - \eref{eq:piecesmod_last} yields the required
second derivatives of the metric,
\numparts
\begin{eqnarray}
\fl \gamma_{ab,cd}Q^aQ^bK^aK^b  =  
	i\sin^2\theta\Bigg\{ \f{8f^{\prime\prime\prime\prime}(u+2r)}{r} 
		- \f{16f^{\prime\prime\prime}(u+2r)}{r^2}
		+\f{4f^{\prime\prime}(u)+20f^{\prime\prime}(u+2r)}{r^3}
	\nonumber\\		
	\qquad\qquad\qquad
		+\f{12[f^{\prime}(u)-f^{\prime}(u+2r)]}{r^4}\Bigg\} \;,
		\\
\fl \gamma_{ab,cd}Q^aK^bQ^cK^d  = 
	i\sin^2\theta\Bigg\{
		\f{4[f^{\prime\prime}(u)-f^{\prime\prime}(u+2r)]}{r^3}
		+\f{12[f^{\prime}(u)+f^{\prime}(u+2r)]}{r^4}		
	\nonumber\\	
	\qquad\qquad\qquad
		+\f{12[f(u)-f(u+2r)]}{r^5}\Bigg\}\;,\\
\fl \gamma_{ab,cd}K^aK^bQ^cQ^d  =  
	i\sin^2\theta\Bigg\{
		\f{4[f^{\prime\prime}(u)-f^{\prime\prime}(u+2r)]}{r^3}
		+\f{12[f^{\prime}(u)+f^{\prime}(u+2r)]}{r^4}
	\nonumber\\
	\qquad\qquad\qquad		
		+\f{12[f(u)-f(u+2r)]}{r^5}\Bigg\} .
\end{eqnarray}
\endnumparts
The subsequent evaluation of  \eref{eq:psi0} leads to
\begin{eqnarray}
\fl \Psi_{0[b]}:=\Psi_0 & = & i\sin^2\theta\Bigg\{ 
	\f{f^{\prime\prime\prime\prime}(u+2r)}{r} 
	-\f{2f^{\prime\prime\prime}(u+2r)}{r^2} 
	+\f{3f^{\prime\prime}(u+2r)}{r^3} 
	-\f{3f^{\prime}(u+2r)}{r^4} 
\nonumber\\
\fl&&\qquad\quad
	-\f{3}{2}\f{f(u)-f(u+2r)}{r^5} \bigg\}\;.
\end{eqnarray}
For the waveform (\ref{eq:C}), we find
\begin{eqnarray}
      \Psi_{0[b]}&=& 0, \quad u > T,  \nonumber \\ 
     \Psi_{0[b]} &=&  \frac{3iCu^2 \sin^2\theta}{4r^5} ,  \quad u <0 , \, u+2r > T .
     \label{psi_m_asympt}
\end{eqnarray}
Consequently, the weak asymptotic stationarity condition \eref{ASC}
implies that $C=0$ and rules out the
B-mode memory \eref{mem_mag_pure} arising from
homogeneous ingoing-outgoing waves.


\subsubsection{Quadrupole E-mode gravitational wave memory}

Ingoing-outgoing waves for electric type homogeneous waves can be generated
from the dual ${}^*K^{abcd}$ of \eref{eq:qmgrwave},
\begin{eqnarray}
\fl {}^*K^{acbd} =  \f{1}{2}\eta^{ab}_{\p{ab}ef}K^{efcd}=
    \frac{1}{3}(\eta^{ab}\eta^{cd}      -\eta^{ad}\eta^{bc})
    +  \f{1}{4}T^{[a} Z^{c]}T^{[b} Z^{d]}
	-\f{1}{4}X^{[a} Y^{c]}X^{[b} Y^{d]} .
      \label{eq:qegrwave}
\end{eqnarray}
The previous procedure for the magnetic type gravitational waves,
using ${}^*K^{abcd}$ instead of $K^{abcd}$, leads to the
quadrupole E-mode gravitational memory
related to the B-mode memory (\ref{mem_mag_pure}) by
\begin{equation}
\Delta \sigma_{[e]} = - i\Delta \sigma_{[b]} = C\sin^2\theta\;
\end{equation}
(also compare with \eref{E_B_sigma}).

Similarily, we find $\Psi_{0[e]} =- i\Psi_{0[b]}$ ,
so, referring to \eref{psi_m_asympt},
\begin{eqnarray}
      \Psi_{0[e]}&=& 0, \quad u > T,  \nonumber \\ 
     \Psi_{0[e]} &=&  \frac{3Cu^2 \sin^2\theta}{4r^5} ,  \quad u <0 , \, u+2r > T .
     \label{psi_e_asympt}
\end{eqnarray}
Consequently, the weak asymptotic stationarity condition \eref{ASC}
again requires $C=0$ and rules out E-mode memory arising from
homogeneous ingoing-outgoing waves. 


\section{Gravitational Memory}
\label{sec:mem}

We now discuss our central issue, the global analysis of the  E and B-mode patterns
of the linearized gravitational  memory. For this purpose,  we consider the general asymptotic
solution of the linearized Einstein equations at $\mc{I}^+$ and, in particular,
the limiting behavior at $u=\pm\infty$, subject to
the asymptotic conditions on the matter and metric discussed in Sec.~\ref{sec:asymp}.

The solution of the Einstein equations, as detailed in Sec.~\ref{sec:bs},
assumes an asymptotic expansion of the matter and metric variables in terms of
a $1/r$ expansion. For the metric variables, we assume, in accord with \eref{metric_gauge},
that the strain variable $J$ has the expansion
\begin{eqnarray}
\label{asypm_J}
J_{} & = & J_{[1]}r^{-1} + J_{[2]}r^{-2} +  J_{[3]}r^{-3} + \mbox{higher order terms}\;, 
\end{eqnarray}
and, for the matter fields, we assume in accord with {  \eref{cond_rho_ab_1}-\eref{fin_int_flux_rhouu_N}}
\begin{equation}
\rho_{rr}  = \f{ \rho_{rr[4]}}{r^{4}}+...\;,\;\; \\
P  =  \f{P_{[3]}}{r^{3}} +...\;,\;\;\\
S_{0}=\f{S_{0[2]}}{r^{2}}+...\;,\;\;\\
S=\f{S_{[2]}}{r^{2}}+...\;.\;\;\\
\end{equation}
We proceed to use these expansions to integrate the linearized equations.

The integration of the  $\beta$-hypersurface equation gives the asymptotic dependence
\begin{equation}
      \beta= -\frac{\kappa \rho_{rr[4]}}{8r^2} + O(1/r^3) \;,
\end{equation}
and together with \eref{asympt_mat_du}
\begin{equation}
\label{du_b2}
\partial_u\beta_{[2]} = 0\;\;,\qquad u=\pm\infty.
\end{equation}
Integration of the $U$-hypersurface equation (\ref{hyp_U_s0}) yields
\numparts
 \begin{eqnarray}
\label{eq:sol_Ue}
 U_{} &=& \f{U_{[2]}}{r^2}+ \f{U_{[3]}}{r^3}+\Big(-\frac{1+3\ln r }{9r^3}\Big)U_{[log]}
     + \mbox{higher order terms}  \; ,\\
 U_{[2]}&=&  -\frac{1}{4} (\eth \bar \eth +2) J_{[1]}  \; ,\label{def_U2}\\
 U_{[3]} &=& -\f{L}{3}\label{def_L} \; ,\\
  U_{[log]} &=&(\eth \bar \eth +2) J_{[2]} -8 \beta_{[2]} +2\kappa P_{[3]} \;,
\end{eqnarray}
 \endnumparts
where the real and imaginary parts of the function of integration $L(u,x^A)$, appearing in $U_{[3]}$
are, respectively, the dipole-moment aspect and the angular-momentum aspect.  
Inserting the  asymptotic solutions for $\beta$ and $U$ into the evolution equation (\ref{ev_s0})
yields
\numparts
\begin{eqnarray}
0&=&\partial_u J_{[2]}\;,\label{du_J2}\\
0&=&U_{[log]} = (\eth \bar \eth +2) J_{[2]}  -8 \beta_{[2]} +2\kappa P_{[3]}\;\;,\label{eq:j2}\\
0&=&2\partial_u J_{[3]} +J_{[2]} +U_{[3]}+2\beta_{[2]} +\kappa S_{[2]}.\label{duJ3_1}
\end{eqnarray}
\endnumparts
As a consequence of (\ref{eq:j2}), the logarithmic term in \eref{eq:sol_Ue} vanishes, i.e.
 \begin{equation}
\label{uexp}
 U_{} = \frac{U_{[2]}}{r^2}  +\frac{U_{[3]}}{r^3}+ \mbox{higher order terms},
 \end{equation}
 which is consistent with asymptotic expansion in $1/r$ required by the peeling property.
Subsequently, integration of the $W$- hypersurface equation (\ref{hyp_W_s0}),
while using \eref{uexp}, gives 
\numparts
\begin{eqnarray}
W &=&  \f{W_{[1]}}{r} +\f{W_{[2]}}{r^2} +...\;\;,\label{W_sol_exp}\\
W_{[1]}&:=& -2M\;\;,\label{def:W_1}\\
W_{[2]}&:=& - \f{1}{2}\bar \eth \eth U_{[e1]}
             - ( \eth\bar \eth+2) \beta_{[2]}+\frac{\kappa}{2}(\eth\bar\eth P_{[e3]}+S_{0[2]})
             \;\;,
             \label{def:W_2}
\end{eqnarray}
\endnumparts
where the function of integration $M=M(u,x^A)$ is the mass aspect, normalized
so that the $uu$-component of the metric perturbation equals $2M/r$ in the static case.
This general solution must also obey the supplementary conditions
at some radius $r$. In the next two sections,
we impose these conditions  in the limit of $\mc{I}^+$, where their real  and
imaginary parts constrain the $E$ and $B$-mode memory, respectively.
The matter terms which appear in the spin-weight 0 version \eref{supp_rr_u_s0}
and \eref{supp_rr_A_s0} are expanded according to  { \eref{cond_rho_ab_1} and \eref{mat_s0_cond} as}
\begin{equation}
\label{supp_asymp_mat}
\rho_{uu} = \f{\rho_{uu[2]}}{r^2} + \f{\rho_{uu[3]}}{r^3} + \f{\rho_{uu[4]}}{r^4} +...\;,\qquad
    N = \f{N_{[2]}}{r^2}+\f{N_{[3]}}{r^3} +  ...\;.
\end{equation}


\subsection{Implications for the E-mode gravitational memory}
\label{sec:mem_E}

The supplementary condition \eref{supp_rr_u_s0}, i.e. $R_{uu}=\kappa\rho_{uu}$,
is purely real, whereas \eref{supp_rr_A_s0}, resulting from the spin-weight 0 version
of $q^AR_{uA} = \kappa \mc{N}$, has both a real and imaginary part. In order
to constrain the E-mode memory we consider the real part
of \eref{supp_rr_A_s0},
\begin{eqnarray}
\fl \qquad 0=
\partial_u \Big[\f{1}{4}(\overline\eth\eth +2)J_{[e]}   - \beta
    -\f{1}{2} r^2 \partial_r U_{[e]}
    \Big]
    +   \f{1}{2} \partial_rW
     +\f{1}{2}\partial^2_r(r^2U_{[e]})
     -\kappa N_{[e]} .
     \label{supp_rr_A_s0_ele}
\end{eqnarray}
Substitution of the asymptotic expansions for $\beta$, $W$, and the real parts
$J_{[e]}$ and  $U_{[e]}$, yields the leading order terms
\numparts
\begin{eqnarray}
\fl\quad[r^{-2}\mbox{ part of } \eref{supp_rr_u_s0}]\;\; 0  =  
     \partial_u W_{[1]} -\partial_u\eth\bar \eth U_{[e2]} -\kappa \rho_{uu[2]} \;, \label{supp1_c1_emem}\\
\fl\quad[r^{-1}\mbox{ part of } \eref{supp_rr_A_s0_ele}]\;\; 0  =   
        \partial_u (\eth \bar\eth +2)J_{[e1]} +4\partial_u  U_{[e2]} \label{supp1_c2_emem} \; ,\\
\fl\quad[r^{-2}\mbox{ part of } \eref{supp_rr_A_s0_ele}] \; \;
	0  =  \partial_u (\eth \bar\eth +2)J_{[e2]}+ 6\partial_u  U_{[e3]} - 4\partial_u \beta_{[2]}  -2 W_{[1]}
	-4\kappa N_{[e2]}.
	\label{supp1_c3_emem}
\end{eqnarray}
\endnumparts

Substitution of the electric part of $U$ from \eref{def_U2} into \eref{supp1_c2_emem} shows
that  \eref{supp1_c2_emem} is already satisfied. 
Using the formulae \eref{def_rad_mem_spin0} and \eref{asyp_strain_calc} for the strain,
along with \eref{def_U2} and  \eref{def:W_1}, the supplementary condition
\eref{supp1_c1_emem} implies
\begin{equation}
\partial_u M = \f{1}{4}\partial_u\eth\bar \eth (\eth \bar \eth +2)\Sigma_{[e]} -\frac{\kappa}{2} \rho_{uu[2]}\;.
\end{equation}
Integration over retarded time from $u=-\infty$ to $u=\infty$ gives
\begin{equation}
\label{emem_1}
\Delta \eth\bar \eth (\eth \bar \eth +2)\Sigma_{[e]}= 4\Delta M + 2\kappa \int_{-\infty}^{+\infty}  \rho_{uu[2]} du.
\end{equation}
Here the contribution from $\rho_{uu[2]}$ is the null memory
arising from the net transport of energy to infinity by the matter.
In the absence of null memory, we thus obtain 
that the E-mode memory $\Delta\Sigma_{[e]}$ is related to the change in the  mass aspect
$\Delta M$ .

Next, the supplementary condition \eref{supp1_c3_emem}, along with
the electric part of \eref{eq:j2},  gives
\begin{equation}
\label{diff_j2_supp_N}
0=2\kappa \partial_u P_{[e3]}- 4\partial_u \beta_{[2]} - 6\partial_u  U_{[e3]} +2W_{[1]} +4\kappa N_{[e2]}.
\end{equation}
 {Differencing   \eref{diff_j2_supp_N}  between $u=\pm \infty$, while employing the
matter conditions  \eref{fin_int_flux_rhouu_N} and \eref{asympt_mat_du}, the condition}
\eref{du_b2}, and the relations \eref{def_L} and \eref{def:W_1}, gives
\begin{equation}
\label{emem_2}
  \Delta \partial_u  L_{[e]}  = 2\Delta  M\;.
\end{equation}
Consequently \eref{emem_1} shows that the electric memory  $\Delta \Sigma_{[e]}$
can alternatively be expressed in terms of the dipole-moment aspect $L_{[e]}$ by
\begin{equation}
\label{emem_3}
\Delta \eth\bar \eth (\eth \bar \eth +2)\Sigma_{[e]} = 2\Delta \partial_u L_{[e]}
      + 2\kappa \int_{-\infty}^{+\infty}  \rho_{uu[2]} du \;.
\end{equation}

Next, application of the weak asymptotic stationarity condition \eref{ASC2}
to the $u$-derivative of \eref{duJ3_1},  { together with \eref{du_b2}, \eref{def_L} and \eref{du_J2},} gives
\begin{equation}
 \Delta \partial_u L_{[e]} = 3{\kappa} \Delta \partial_u S_{[e2]}.
\end{equation}
As a result, \eref{emem_3}  relates the E-mode memory to the net change in the
$u$-derivative of the anisotropic stress according to
\begin{equation}
\label{elec_mem}
    \Delta \eth\bar \eth (\eth \bar \eth +2)\Sigma_{[e]} = 6\kappa \Delta \partial_u S_{[e2]}\
      + 2\kappa \int_{-\infty}^{+\infty}  \rho_{uu[2]} du \;\;.
\end{equation}
Thus, except for the null memory arising from $ \rho_{uu[2]}$, the only memory effect
allowed by the weak asymptotic stationarity condition must arise from a
matter distribution with the asymptotic behavior $\Delta \partial_u S_{[e2]}\ne 0$.
A  detailed discussion of
this possibility is postponed to Sec.~\ref{sec:mem_matter}.


\subsection{Implications for the  B-mode gravitational memory}
\label{sec:mem_B}

As seen from (\ref{rad_memory}), the B-mode gravitational memory
$\Delta\Sigma_{[b]}$ is determined by the imaginary part  $J_{[b1]}$.
As only the fields $J$, $U$, $P$, $N$ and $S$ have imaginary parts, 
the only relevant supplementary equation is the imaginary part of
\eref{supp_rr_A_s0},
\begin{eqnarray}
\fl \quad 0 =\partial_u \Big[\f{1}{4}(\overline\eth\eth +2)J_{[b]}   -\f{1}{2} r^2 \partial_r U_{[b]} \Big]
+
     \f{1}{2}\partial^2_r(r^2U_{[b]})
     +\f{1}{2}\overline \eth \eth U_{[b]}
     -\kappa N_{[b]} \;.
     \label{supp_rr_A_s0_mag}
\end{eqnarray}
The leading two orders of \eref{supp_rr_A_s0_mag}, along with the imaginary parts of
\eref{asypm_J}, \eref{uexp} and \eref{supp_asymp_mat},  give
\numparts
\begin{eqnarray}
(r^{-1}):\;\; 0=\f{1}{4}\partial_u(\overline\eth\eth +2)J_{[b1]} +\partial_u U_{[b2]} \label{supp1_c1_Bmem}\; , \\
(r^{-2}):\;\; 0=\f{1}{4}\partial_u(\overline\eth\eth +2)J_{[b2]} 
   +\f{3}{2}\partial_u U_{[b3]}+\f{1}{2}\overline \eth \eth U_{[b2]}  -\kappa N_{[b2]}\label{supp1_c2_Bmem}.
\end{eqnarray}
\endnumparts
The imaginary  part of \eref{def_U2} implies that  \eref{supp1_c1_Bmem} is already satisfied.
From the imaginary parts of   {\eref{fin_int_flux_rhouu_N}, \eref{def_U2}, \eref{def_L} and \eref{du_J2}  }
the second condition \eref{supp1_c2_Bmem} simplifies at $u=\pm\infty$ to 
\begin{equation}
0=\f{1}{4}\overline \eth \eth (\overline\eth\eth +2)J_{[b1]} +\partial_u L_{[b]}\; ,
\end{equation}
so that
\begin{equation}
\Delta \overline \eth \eth (\overline\eth\eth +2)\Sigma_{[b]} = -2\Delta \partial_u L_{[b]}\;.
\end{equation}

Thus the B-mode memory $\Delta \Sigma_{[b]}$ is related to the angular momentum aspect $L_{[b]}$
in a similar way that the E-mode memory is related to the dipole-moment aspect $L_{[e]}$. Also note, there
is no analogue of the null memory for the B-mode case.
As in the E-mode case, the B-mode memory can be related to the asymptotic anisotropic stress $S$
by applying the asymptotic stationarity condition \eref{ASC2} to the magnetic part of \eref{duJ3_1}  {together with \eref{def_L} and \eref{du_J2}},
\begin{equation}
\label{mag_mem}
\Delta \overline \eth \eth (\overline\eth\eth +2)\Sigma_{[b]} = -6\kappa \Delta \partial_u S_{[b2]} \;.
\end{equation}

Thus, after using weak asymptotic stationarity to rule out the known source of B-mode memory, i.e. a
homogeneous wave,  a non-compact 
matter distribution with the asymptotic stress $\Delta \partial_u S_{[b2]}\ne 0$
remains the only possibility.


\subsection{Matter implications for E and B-mode gravitational memory}
\label{sec:mem_matter}

From \eref{elec_mem} and \eref{mag_mem}, we have shown that
the weak asymptotic stationarity condition condition leads to
\begin{equation}
\label{all_mem}
   \Delta \overline \eth \eth (\overline\eth\eth +2)\Sigma = 
   6\kappa \Delta \partial_u \Big(S_{[e2]}-iS_{[b2]}\Big)
   + 2\kappa \int_{-\infty}^{+\infty}  \rho_{uu[2]} du 
\end{equation}
In the absence of null memory $\int_{-\infty}^{+\infty}  \rho_{uu[2]} du =0$
so that \eref{all_mem} reduces to
\begin{equation}
\label{mem_no_null}
   \Delta \overline \eth \eth (\overline\eth\eth +2)\Sigma = 
   6\kappa \Delta \partial_u \Big(S_{[e2]}-iS_{[b2]}\Big)  \; ,
\end{equation}
which shows that all remaining possible sources of gravitational memory are determined by
the retarded time derivatives of the stress $\partial_uS_{[2]}$
at $u=\pm\infty$. Recall that $\Sigma$ contains no
monopole or dipole harmonics which otherwise would be allowed by the inverse of the
$ \overline \eth \eth (\overline\eth\eth +2)$ operator.
{Since the matter stresses $S$ were defined in Sec. \ref{sec:bs} with respect to a most general energy momentum tensor and without specification of the type of matter,} we now discuss the asymptotic behavior of  $\partial_uS_{[2]}$ for various sources. 

 { \it Sources of compact support:} For matter which vanishes in a neighborhood
  of ${\mathcal I}^+$ at all finite $u$, it immediately follows that
  $\partial_u S_{[2]}=0$ for all $u$. Note that this
  does not rule out matter confined inside an expanding (or contracting) worldtube $r=R(u)$, where 
  $R(u)\rightarrow \infty$ as $u\rightarrow \pm \infty$.
   
 {\it Scalar fields:} The stress-energy tensor of
    a scalar field  $\Phi$ with mass $m$,
    \begin{equation}
           T_{ab} = (\partial_a\Phi)(\partial_b\Phi)-\f{1}{2}\eta_{ab}\Big[\eta^{cd}(\partial_c\Phi)(\partial_d\Phi)
            + \f{m^2}{\hbar^2}\Phi^2\Big],
     \end{equation}
    gives rise to the anisotropic stress
    \begin{equation}
          \eth^2 S = (\eth \Phi)^2 .
    \end{equation}
   For a massive scalar field on a Minkowski background,  it has been shown that $\Phi$
    falls off  faster than any finite power $1/r^n$ at $\mc{I}^+$~\cite{massiveScri,massiveScri2}.
Therefore $S_2$ vanishes for all $u$. 
     
For a massless scalar field, the asymptotic radiative behavior 
	 $\Phi(x^a) = \Phi_{[1]}(u,x^A)r^{-1}+O(r^{-2})$ implies
\begin{eqnarray}
               T_u^r & =- &\f{1}{r^2} (\partial_u \Phi_{[1]})^2 
           +...\quad,\qquad 
                  \eth^2 \partial_u S_{[2]}= 2 (\eth \Phi_{[1]})\eth \partial_u \Phi_{[1]}.
                   \label{S_2_Phi}
\end{eqnarray}
The requirement that the $u$-integration of the energy flux $T_u^r$  to $\mc{I}^+$
be finite implies that  $\partial_u\Phi_{[1]} = 0$ for $u=\pm\infty$. 
Thus $\Delta \partial_u S_{[2]} = 0$.
 
 {\it Electromagnetic fields:}  
  	  The stress-energy tensor of an electromagnetic field $F_{ab}$, 
	  \begin{equation}
	\label{T_em}
	T_{ab} = F_{ac}F_b^c-\f{1}{4}\eta_{ab}F^{cd}F_{cd},
	\end{equation}
	gives rise to an anisotropic stress
   \begin{equation}
          \eth^2 S = q^A q^B  F_{Ac}F_B^c.
    \end{equation}
    The asymptotic behavior of the electromagnetic field, dictated by the peeling property
    implies (in null spherical coordinates) 
    $q^AF_{A u}=O(1)$, $q^A \bar q^B F_{AB} =  O(1)$,
     $F_{ur}=O(1/r^2)$ and $q^A F_{Ar} = O(1/r^2)$.
    As a result, 
    \begin{eqnarray}
         T_u^u&=&  \frac{1}{r^2} |q^AF_{Au[0]}|^2 +...\;,\quad
	  \eth^2 S_{[2]} = -2q^A q^B F_{Au[0]}F_{Br[2]} \; .
    \end{eqnarray}
    As in the scalar case, the requirement that the $u$-integration of the energy flux $T_u^r$  to $\mc{I}^+$
    be finite implies that  $q^AF_{Au[0]}= 0$ for $u=\pm\infty$. 
    Thus $S_{[2]} = 0$ at $u=\pm \infty$ and consequently $\Delta \partial_u S_{[2]} = 0$.
  
 {\it Perfect fluid halo:} The stress-energy tensor of a perfect fluid, with  matter density $\rho$,
  pressure $p$ and 4-velocity $v^a$,
 \begin{equation}
	T_{ab} = (\rho+p)v_a v_b +p\eta_{ab} ,
  \end{equation}
gives rise to the anisotropic stress
\begin{equation}
       \eth^2 S = (\rho+p) (q^A v_A)^2 = r^4 (\rho+p) (q_A v^A)^2 .
\end{equation}
Here, the normalization $v_a v^a=-1$ requires $v^u|_{r=\infty}\neq 0$, $v^r=O(1)$ and $v^A =O(1/r)$.
The finite energy-momentum conditions,   $T_u^u=O(1/r^4)$ and $T_r^u=O(1/r^4)$,  imply
$\rho=O(1/r^4)$ and $p=O(1/r^4)$ and consequently
\begin{equation}
 \eth^2  \partial_u  S_{[2]} = q_A q_B \partial_u\bigg [(\rho_{[4]}+p_{[4]})  v_{[1]}^A v_{[1]}^B \bigg ]\;\;.
\end{equation}
The leading terms in the $A$-component of the conservation laws (\ref{cons_law_T}) imply
at $u=\pm \infty$ that
\begin{equation}
    0=    \partial_u \bigg[ (\rho_{[4]}+p_{[4]}) v_{[1]}^A v^u_{[0]} \bigg ] . 
    \label{eq:Adot}
\end{equation}
The equations of motion $(\kron{a}{b}+v^a v_b)\nabla_cT^{cb}=0$ 
give to leading order in $1/r$
\begin{equation}
    (\rho_{[4]} +p_{[4]}) \partial_u v^u_{[0]} + v^u_{[0]} \partial_u p_{[4]} =0\;,
    \label{eq:umot}
\end{equation} 
\begin{equation}
    (\rho_{[4]} +p_{[4]}) v^u_{[0]} \partial_u v^r_{[0]} +(-1 + v^r_{[0]} v^u_{[0]}) \partial_u p_{[4]} =0\;,
\end{equation} 
\begin{equation}
    (\rho_{[4]} +p_{[4]}) \partial_u v^A_{[1]} +v^A_{[1]}  \partial_u p_{[4]} =0
    \label{eq:Amot}.
\end{equation} 
We have
\begin{eqnarray}
\fl         \eth^2v^u_{[0]}\partial_u S_{[2]} 
	&=&  
		q_A q_B v^u_{[0]} \partial_u\bigg [(\rho_{[4]}+p_{[4]})  v_{[1]}^A v_{[1]}^B \bigg ]
       \nonumber \\
      & =&- q_A q_B\Bigg\{ v^u_{[0]} (\partial_u v_{[1]}^A)(\rho_{[4]}+p_{[4]}) v_{[1]}^B 
          +v^A \partial_u \Big[ (\rho_{[4]}+p_{[4]}) v_{[1]}^B v^u_{[0]} \Big ]
          \nonumber \\
          &&\qquad \quad- (\partial_u v_{[0]}^u)(\rho_{[4]}+p_{[4]})  v_{[1]}^A v_{[1]}^B \Bigg\}.
 \end{eqnarray}
As a result, a direct application of (\ref{eq:Adot}), (\ref{eq:umot}), and (\ref{eq:Amot}) gives,
after some algebra,
\begin{equation}
     \eth^2v_{[0]}^u\partial_u S_{[2]} =0, \quad u=\pm \infty.
\end{equation}
Therefore $\Delta \partial_u S_2=0$ for a perfect fluid halo.


\section{Summary}
\label{sec:concl}

In the context of linearized theory,
we have shown that the weak asymptotic stationarity condition, along with standard asymptotic
conditions on the metric and matter stress-energy tensor, imply that
the only possible sources of gravitational radiation memory, other than null memory,
require matter with asymptotic stress satisfying $\Delta \partial_u S_2\ne 0$.
While, on the basis of general principles,
we cannot rule out such an anisotropic, time dependent stress, we have shown that it is ruled out
by common sources including matter confined to an expanding or contracting worldtube, scalar
fields, electromagnetic fields and a perfect fluid halo. Barring exotic matter sources, our results
imply that the E-mode memory effect is restricted to null memory and the two known sources
that are eliminated by the weak asymptotic stationarity condition, i.e. burst memory
(and its time reversed counterpart) and homogeneous wave memory. The results for B-mode
memory are more restrictive since there is no B-mode analogue to null memory or burst memory.
Thus the only known source of B-mode memory appears to be of primordial origin, corresponding in the linearized
theory to  a source free, ingoing-outgoing gravitational wave entering from past null infinity. 

Although our results are based upon linearized theory, it is straightforward to extend
the underlying approach, an asymptotic expansion of the Bondi-Sachs metric 
along the outgoing null cones, to the nonlinear theory. Thus our results set the stage for
an investigation of how the presence of black holes affect the sky pattern of the memory effect.

 
\ack
T.M is grateful for hospitality of the AEI in {Golm} where this project was initiated and
appreciates valuable discussions and support  from  P. Jofr\'e, Q. Kral, A. Bonsor, C. Malone and A. J. Penner.  
J.W. was supported by NSF grant PHY-1505965 to the University of Pittsburgh.

\appendix


\section{Global electromagnetic  memory  revisited}
\label{app:em}

Here we revisit  the global electromagnetic memory analysis and  show that a
weaker asymptotic stationarity condition than the one employed in~\cite{globalemmem} yields
the same results. With a slight change of notation from~\cite{globalemmem},
we adopt the spin-weight formulation analogous to the linearized gravitational case. 
In this notation, the Maxwell field and vector potential, $F_{ab}= 2\partial_{[a}A_{b]}$, have angular
components in null spherical coordinates represented by the spin-weight 0 potentials $\msc{E}$
and $\msc{A}$ according to
\begin{equation}
  \eth \msc{E} =q^B F_{Bu}, \quad  \eth \msc{A}=q^BA_B ,
\end{equation}
where, by convention, $\msc{E}$ and $\msc{A}$ have no $l=0$ component.
We adopt a null gauge $A_r=0$, which is the electromagnetic analogue
of the Bondi-Sachs gauge~\cite{globalemmem, tam}.
We use the remaining gauge freedom,
$A_a\rightarrow A_a +\partial_a \Lambda(x^a)$, to set
 $A_{u[0]}= A_u |_{{\mathcal I}^+}=0$.
The remnant gauge freedom 
$A_B \rightarrow A_B + \partial_B \, \Lambda(x^C)$
is the electromagnetic analogue of a
BMS supertranslation.  
The electromagnetic radiation memory for a test particle with unit charge and mass is
\begin{equation}
 \Delta\eth \msc{V}: = \lim_{r\rightarrow \infty} \int_{-\infty}^\infty \eth \msc{E}  d u 
    =  - \Delta  \eth \msc{A}_{[0]}\; .
 \label{em_mem_s0}
\end{equation}

The peeling property of an isolated system requires that the electromagnetic Newman-Penrose
component $\Phi_0:=\f{1}{2}F_{ab}K^a Q^b=O(1/r^3)$. The {\it new}
weak asymptotic stationarity condition is
\begin{equation}
     \Delta \lim_{r\rightarrow\infty}r^3   \partial _u \Phi_0
    =\Delta  \partial _u\Phi_{0[3]}
    = - \Delta \partial_u \eth  \msc{A}_{[1]}
     =0\;.
      \label{astat_em_mem}
\end{equation}

There are three known causes of electromagnetic memory: burst memory due to the ejection of
charged particles, homogeneous wave memory due to source free waves and null
memory due to the flow of charge  to ${\mathcal I}^+$ by a (hypothetical) charged massless field
or fluid. We now show that the weak asymptotic stationarity condition (\ref{astat_em_mem})
rules out burst memory and homogeneous wave memory.

Burst memory results from the asymptotic behavior of a boosted Coulomb field, which
can be described in the Lorentz covariant form
\begin{equation}
A_a = \f{q}{R}v_a\;\;,\qquad
F_{ab} = \f{q}{R^3}(x_a v_b - x_b v_a)\;,
\end{equation}
where $v_a$ the four velocity of a particle with charge $q$ and $R$ is given by \eref{R_Ra}.
A charged particle which is initially at rest and ejected with
velocity $V$ in the $z$-direction gives rise to the non-zero burst memory
\begin{equation}
   \eth \Delta \msc{V}= \frac {q V \sin\theta}{1-V \cos\theta}.
\end{equation}
The resulting electromagnetic Newman-Penrose scalar is
\begin{equation}
       \Phi_{0} =\f{q}{2R^3}(K^ax_a)(Q^bv_b) 
     = \f{q (1-V^2) V\sin\theta}{2(1-V\cos\theta)^3}\f{u}{r^3} +O(r^{-4}) 
\end{equation}
so that the weak asymptotic stationarity condition \eref{astat_em_mem} rules 
out burst memory. 

Homogeneous electromagnetic wave memory can be
described using an antisymmetric Hertz potential
$H^{ab}=H^{[ab]}$. The vector potential
$ A^a=\partial_b H^{ab}$
then satisfies the Lorentz gauge condition $\partial_b A^b=0$ and generates a solution of
Maxwell's equations provided $H^{ab}$ satisfies the wave
equation $\eta^{ab}\partial_a\partial_b H^{cd}=0$.
Following \cite{globalemmem}, the  purely E-mode Hertz potential 
\begin{equation}\label{Hz_ele}
        H^{ab}_{(E)} := (T^a Z^b - Z^a T^b) \frac{f(u) -f(u+2r)}{r},  
\end{equation}
\begin{eqnarray} \label{eq:F}
       f(\tau) = \left\{
     \begin{array}{ccc}
     C \tau& : &T<\tau \\
      F(\tau)& : &0\le \tau \le T \\
      0& : &\tau<0  \; ,
\end{array}\right.
\end{eqnarray}
where $F(\tau)$ is chosen to make the solution smooth,
yields the non-zero E-mode memory
$$\Delta \msc{V}_{[e]}=- C\cos\theta .$$
The corresponding Newman-Penrose component satisfies
\begin{eqnarray}
  \fl    \Phi_{[e]0} &=& 0, \quad u >T, \;\mbox{and }\;\;   
      \Phi_{[e]0} =-\f{1}{2}\frac{Cu}{r^3}\sin\theta\, , \quad u < 0, \, u+2r >T,
      \label{Phi0_e_asympt}
\end{eqnarray} 
so that
\begin{equation}
  \Delta  \Phi_{[e]0} =\frac{1}{2}\frac{Cu}{r^3}\sin\theta.
\end{equation}
Thus the weak asymptotic stationarity condition
\eref{astat_em_mem} rules out homogeneous wave E-mode memory.

Homogeneous B-mode solutions of the Maxwell equation are generated
from the dual $H^{ab}_{(B)} ={}^*H^{ab}_{(E)}$ of \eref{Hz_ele}.
The resulting vector potential is purely magnetic with
B-mode memory and B-mode Newman-Penrose
component
\begin{equation}
\Delta\msc{V}_{[b]} = i\Delta\msc{V}_{[e]}\;\;\mbox{ and }\;\;
  \Phi_{[b]0} = -i\Phi_{[e]0}\;.
\end{equation}
Consequently the weak asymptotic stationarity condition \eref{astat_em_mem} also
rules out homogeneous wave B-mode memory.

We now show that the weak asymptotic stationarity condition
rules out all memory sources except E-mode null memory. For that purpose,
we consider Maxwell's equations with 4-current $J^a$, whose angular components
are represented by a spin-weight 0 potential according to $ \eth \msc{J} = q_A J^A $.
In the null gauge, $A_r=0$, Maxwell equations decompose into
the hypersurface, evolution
and supplementary  equations given by, respectively,
\begin{eqnarray}
4\pi r^2 J^u& = & \partial_r\Big( r^2\partial_rA_u - \bar \eth \eth \msc{A}_{[e]}\Big) \;\;,
\end{eqnarray}
\begin{eqnarray}\label{ev_s0_em}
4\pi r^4 \msc{J} & = & 
	r^2\partial_r \Big(2\partial_u \msc{A} -  A_u - \partial_r \msc{A}\Big)
	+i\bar \eth \eth \msc{A}_{[b]}\
\end{eqnarray}
\begin{eqnarray}\label{supp_s0_em}
4\pi r^2 J^r &=&
	\partial_r (\bar \eth \eth \msc{A}_{[e]}) - \partial_u(\bar \eth \eth \msc{A}_{[e]}) + \bar\eth \eth A_u
	-r^2\partial_r\partial_u A_u\;\;.
\end{eqnarray}

As discussed in \cite{globalemmem}, if the current density is modeled after a charged fluid then
the finiteness of its total angular momentum requires in Cartesian coordinates
 $x^i J^k -x^k J^i =O(r^{-3-\epsilon}), \; \epsilon>0$,
which implies in null spherical coordinates that $J^A=O(r^{-5-\epsilon})$. In addition,
the finite flux of charge to ${\mathcal I}^+$ requires $J^r=O(r^{-2})$ with $J^r_{[2]}|_{\pm \infty}=0$.
Thus the current satisfies the asymptotic conditions
\begin{equation}
       \Delta J^r_{[2]}  = \Delta  \msc{J}_{[4]}=0\;.
         \label{J_asymp_s0_cond}
\end{equation}

{\it Global E-mode memory:}
As discussed in \cite{globalemmem}, the electromagnetic E-mode memory
is governed by the supplementary equation \eref{supp_s0_em} evaluated at ${\mathcal I}^+$,
\begin{equation}\label{supp_em_scri}
   \{ 4\pi r^2  J^r +\partial_u \eth \bar \eth \msc{A}_{[e]}  +r^2 \partial_u  \partial_r A_u \}|_{{\mathcal I}^+}=0.
\end{equation}
Integration of \eref{supp_em_scri} shows that the radiation memory (\ref{em_mem_s0})
is governed by two terms,
\begin{equation}
    \bar \eth\eth( \Delta  \msc{V}_{[e]})  = -\Delta A_{u[1]}
          + 4\pi \int_{-\infty}^{\infty} J^r_{[2]}  du \; ,
    \label{eq:mem1_s0}
\end{equation}
where the second term is the null memory  {and \eref{eq:mem1_s0} is the electromagnetic counterpart to the E mode of the linearized gravitational wave memory  \eref{elec_mem}.}

The real part of the evolution equation \eref{ev_s0_em}
gives at $\mc{I}^{+}$ 
  \begin{eqnarray}
4\pi \msc{J}_{[e4]} & = & 
	-2\partial_u \msc{A}_{[e1]} +  A_{u[1]}\;.
\end{eqnarray}
Weak asymptotic stationarity \eref{astat_em_mem} along with the current
condition \eref{J_asymp_s0_cond} imply that
$ \Delta A_{u[1]}=0$. Consequently, \eref{eq:mem1_s0} 
shows that weak asymptotic stationarity eliminates all E-mode
memory except null memory.

{\it Global B-mode memory:}
The B-mode component of the electromagnetic memory (\ref{em_mem_s0}),
 \begin{equation}
	\label{mag_mem_app}
	\Delta \msc{V}_{[b]}= -\Delta \msc{A}_{[b0]} ,
 \end{equation}
is governed by the imaginary part of the evolution equation  \eref{ev_s0_em},
 \begin{eqnarray}
    0 & = & r^2\partial_r \Big(4\partial_u \msc{A}_{[b]}  - \partial_r \msc{A}_{[b]} \Big)
	-\bar \eth \eth \msc{A}_{[b]} -4\pi r^4 \msc{J}_{[b]} .
\end{eqnarray}
Its evaluation at $\mc{I}^+$, along with the imaginary part of
the weak asymptotic stationarity condition \eref{astat_em_mem} and
the  current condition \eref{J_asymp_s0_cond}, yields
\begin{equation}
0=	 \Delta \bar \eth \eth \msc{A}_{[b0]} .
\end{equation} 
Thus weak asymptotic stationarity eliminates all B-mode electromagnetic  radiation memory.

\section{Christoffel symbols}\label{app:christ}
Here we list useful Christoffel symbols for the calculation of the curvature quantities.
For simplicity, we define $U_A=q_{AB}U^B$.
\numparts
\begin{eqnarray}
\chr{u}{uu} & = & {2\partial_u\beta -\f{1}{2} \partial_rW -\partial_r\beta}
 \\
\chr{u}{ur} & = & 0 \\
\chr{u}{uA} & = & D_A\beta-\f{1}{2}\partial_r\Big(r^2U_A\Big) \\
\chr{u}{rr} & = & 0 \\
\chr{u}{rA} & = & 0 \\
\chr{u}{AB} & = & rq_{AB}-2r\beta q_{AB}+\f{1}{2}\partial_r(r^2 J_{AB})
\end{eqnarray}
\endnumparts
\numparts
\begin{eqnarray}
\chr{r}{uu} & = & {\f{1}{2} \partial_uW	 -\partial_u\beta+ \f{1}{2} \partial_rW		+\partial_r\beta}
        \\        
\chr{r}{ur} & = & {    \f{1}{2} \partial_rW		+\partial_r\beta}
       \\
\chr{r}{uA} & = & 
      {\f{1}{2}D_AW
     +\f{1}{2}\partial_r(r^2U_A)}
     \\
\chr{r}{rr} & = & 2\partial_r\beta\\
\chr{r}{rA} & = &
   	\f{1}{2}r^2\partial_rU_A
	+D_A\beta\\
\chr{r}{AB} & = & { -rq_{AB} 
      -r(W-2\beta) q_{AB}
                +   r^2D_{(A}U_{B)}
      +\f{1}{2}r^2\partial_uJ_{AB}}  
      \nonumber\\&&
      -\f{1}{2}\partial_r(r^2 J_{AB})
\end{eqnarray}
\endnumparts
\numparts
\begin{eqnarray}
\chr{A}{uu} & = &{ -\partial_uU^A + \f{1}{r^2}D^A\Big(\beta+\f{1}{2}W\Big)}\\
\chr{A}{ur} & = &-\f{1}{2r^2}\partial_r\Big(r^2U^A\Big)+\f{1}{r^2}D^A\beta \\
\chr{A}{uB} & = & 
-\f{1}{2}q^{AC}\Big[D_BU_C-D_CU_B\Big]+\f{1}{2}q^{AC}\partial_u J_{BC} 
\\
\chr{A}{rr} & = & 0 \\
\chr{A}{rB} & = &  \f{1}{r}\kron{A}{B} +\f{1}{2}q^{AC}\partial_rJ_{CB} \\
\chr{A}{BC} & = &
^{(q)}\chr{A}{BC}
+rU^{A} q_{BC} { +D_{(B}J_{C)}^{\p{C}A}-\f{1}{2}D^A J_{BC}}
\end{eqnarray}
\endnumparts


\end{document}